\documentclass[11pt]{article}
\usepackage{latexsym}
\usepackage{amssymb}
\usepackage[dvips]{graphicx}
\usepackage{epsfig}
\usepackage{rotating}
\usepackage{amsfonts}
\usepackage{amsmath}
\begin{document}

\begin{titlepage}
\begin{flushright}
NITheP-09-04\\
ICMPA-MPA/2009/19\\
\end{flushright}

\begin{center}

{\Large\bf Harmonic oscillator in a background
magnetic field in noncommutative quantum phase-space}

Joseph Ben Geloun$^{a,b,c,*}$, Sunandan Gangopadhyay$^{a,\dag}$ and Frederik G Scholtz$^{a,\ddag}$

$^{a}${\em National Institute for Theoretical Physics}\\
{\em Private Bag X1, Matieland 7602, South Africa}\\
$^{b}${\em International Chair of Mathematical Physics
and Applications}\\
{\em ICMPA--UNESCO Chair 072 B.P. 50  Cotonou, Republic of Benin}\\
$^{c}${\em D\'epartement de Math\'ematiques et Informatique}\\
{\em  Facult\'e des Sciences et Techniques, Universit\'e Cheikh Anta Diop, Senegal}

E-mail:  $^{*}$bengeloun@sun.ac.za,\quad $^{\dag}$sunandan@sun.ac.za,
sunandan.gangopadhyay@gmail.com\quad $^{\ddag}$fgs@sun.ac.za  

\today

\begin{abstract}
\noindent We solve explicitly the two-dimensional harmonic oscillator and the harmonic oscillator in a background magnetic
field in noncommutative phase-space without making use of any type of representation. 
A key observation that we make is that for a specific choice of the noncommutative parameters, the time reversal symmetry of the systems get restored since the energy spectrum becomes degenerate. This is in contrast to the noncommutative  
configuration space where the time reversal symmetry of the 
harmonic oscillator is always broken.
\end{abstract}

\end{center}

Pacs : 11.10.Nx

\end{titlepage}

\setcounter{footnote}{0}

\noindent In the last few years, theories in noncommutative 
space have been extensively studied \cite{doug}-\cite{giri}.
The motivation for this kind of investigation 
is that the effects of noncommutativity of space may appear
in the very tiny string scale or at very high energies \cite{dop}.
In particular, two-dimensional noncommutative harmonic oscillators 
have attracted a great deal of attention in the literature \cite{nair}-\cite{mitra}, \cite{sj7}-\cite{giri}. An interesting  observation is that the introduction of noncommutative spatial 
coordinates breaks the time reversal symmetry since the angular momentum states with eigenvalue $-m$ and $+m$ do not have the same energy.

\noindent A description of two-dimensional noncommutative phase-space has been given in the literature \cite{jian, kang, giri} to study the noncommutative harmonic oscillator and noncommutative Lorentz transformations. 
The generalized Bopp shift in this new formulation connecting the
noncommutative variables to commutative variables has also been written down. Using this, the noncommutative harmonic oscillator Hamiltonian has been mapped to an equivalent commutative Hamiltonian and solved
explicitly. A disadvantage of this approach is that one lacks a clear
physical interpretation of the commutative variables that are introduced
in the Bopp shift and subsequently the interpretation of the quantum theory
is obscured. Furthermore, more complicated potentials may lead to highly
non-local Hamiltonians once the Bopp shift is performed. In \cite{sj7},
an algebraic approach was used to solve the noncommutative harmonic
oscillator and within this framework an unambiguous physical interpretation
was also given. With this background it is natural to ask if it is 
possible to solve the spectrum of the harmonic oscillator
on noncommutative phase-space without invoking the Bopp-shift.

\noindent In this letter, we solve the two-dimensional harmonic
oscillator and the harmonic oscillator in a background
magnetic field in  noncommutative phase-space 
working with noncommutative coordinates and momenta without making use of any representation. Our results are in conformity with the ones
existing in the literature. We also show that there exists some interesting
choices of the noncommutative parameters for which the time reversal symmetry can be restored.

\noindent Let us start by considering a two-dimensional 
noncommutative space defined by the following commutation relation 
\begin{equation}
[\hat{x}, \hat{y} ]=i\,\theta
\label{ncb}
\end{equation}
where $\theta$ is the noncommutative parameter.

\noindent It is useful to introduce the complex coordinates
\begin{eqnarray}
\hat{z}= \hat{x}+ i \,\hat{y}, 
\quad \hat{\bar{z}}=\hat{x}- i \,\hat{y}
\end{eqnarray}
so that one can infer from (\ref{ncb})
\begin{eqnarray}
[\hat{z},  \hat{\bar{z}}]= 2\, \theta.
\label{ferm}
\end{eqnarray}
Next, we can introduce the pair of boson annihilation and creation operators $b=(1/\sqrt{2\theta})\hat{z}$ 
and $b^\dagger=(1/\sqrt{2\theta}) \hat{\bar{z}}$ satisfying
the Heisenberg-Fock algebra $[b,b^\dag]=\,1\!\!1$. Therefore, 
the noncommutative configuration space ${\cal H}_c$ is itself a Hilbert space
isomorphic to the boson Fock space ${\cal H}_c = {\rm span}\{|n\rangle, n\in \mathbb{N}\}$, with $|n\rangle = (1/\sqrt{n!})(b^\dag)^{n}|0\rangle$.  
 
\noindent The Hilbert space at the quantum level, denoted by ${\cal H}_q$, 
is defined to be the space of Hilbert-Schmidt operators on ${\cal H}_c$ \cite{hol}
\begin{equation}
\label{qhil}      
\mathcal{H}_q = \left\{ \psi(\hat{z},\hat{\bar z}): \psi(\hat{z},\hat{\bar z})\in \mathcal{B}\left(\mathcal{H}_c\right),\;{\rm tr_c}(\psi(\hat{z},\hat{\bar z})^\dagger\psi (\hat{z},\hat{\bar z}) )< \infty \right\}
\end{equation}
where ${\rm tr}_c$ stands for the trace over ${\cal H}_c$ and
$\mathcal{B}\left(\mathcal{H}_c\right)$ is the set of bounded operators
on $\mathcal{H}_c$.

\noindent With this formalism at our disposal, we now consider the
Hamiltonian $\hat{H}$ of the two-dimensional harmonic oscillator
\begin{eqnarray} 
\hat H=\frac{1}{2m}
\hat{P}_{i}^2 + \frac{1}{2}m\omega^2
\hat{X_i}^2\;,\;(i=1,2)\nonumber\\
 \hat  X_1=\hat X\;,\; \hat X_2=\hat Y\;,\; 
\hat{P}_{1} = \hat{P}_{X}\;,\; \hat{P}_{2} = \hat{P}_{Y}
\label{twohar}
\end{eqnarray} 
on the noncommutative phase-space 
\begin{eqnarray}
[\hat{X}, \hat{Y}]=i\theta\quad,
\quad [\hat{X}, \hat{P}_{X}]=i\tilde{\hbar}=[\hat{Y}, \hat{P}_{Y}]
\quad,\quad[\hat{P}_{X}, \hat{P}_{Y}]=-i\bar\theta
\label{alg1ab1}
\end{eqnarray}  
where capital letters refer to quantum operators 
over ${\cal H}_q$ and $\tilde \hbar$ is the deformed Planck constant, namely
$\tilde \hbar = \hbar + f(\theta,\bar\theta, \hbar)$.
Introducing the complex operators $\hat{Z} = \hat{X}+i \hat{Y}$,
$\hat{\bar Z} = \hat{X}-i \hat{Y}$, $\hat{P}_{Z}= \hat{P}_{X}-i\hat{P}_{Y}$
and $\hat{P}_{\bar Z}= \hat{P}_{X}+i\hat{P}_{Y}$, the algebra (\ref{alg1ab1})
can be rewritten as
\begin{eqnarray} 
[\hat{Z},  \hat{\bar Z} ] = 2\theta\qquad, \qquad
[\hat{Z}, \hat{P}_{Z} ] =  2i\,\tilde{\hbar}=
[\hat{\bar Z}, \hat{P}_{\bar Z} ]\qquad,  \qquad
[\hat{P}_{Z}, \hat{P}_{\bar Z} ]= 2 \bar\theta~.
\label{qbalg}
\end{eqnarray} 
The Hamiltonian $\hat{H}$ of the harmonic oscillator (\ref{twohar}) in 
terms of the complex coordinates reads
\begin{eqnarray}
\hat{H}&=& \frac{1}{4m}(\hat{P}_{Z}\hat{P}_{\bar{Z}}+\hat{P}_{\bar{Z}}\hat{P}_{Z})
+\frac{1}{4}m \omega^2 (\hat{Z}\hat{\bar{Z}}+\hat{\bar{Z}}\hat{Z}).
\label{harham}
\end{eqnarray}
We now rewrite $\hat{H}$ in the matrix form 
\begin{eqnarray}
\hat{H}= \frac{1}{4 m}\; \mathfrak{Z}^\dag\, \mathfrak{Z}\quad,
\quad \mathfrak{Z}= (\hat{Z}',\hat{\bar{Z}}',\hat{P}_{Z},\hat{P}_{\bar{Z}})^t, \;
\hat{Z}'= m\omega \hat{Z},\;\;
\hat{\bar{Z}}' =m\omega  \hat{\bar{Z}}
\label{hamat}
\end{eqnarray}
where the symbol $t$ means the transpose operation. The next objective
is the diagonalization of this Hamiltonian  such that
\begin{eqnarray}
\hat{H}= \frac{1}{4 m} 
A^\dag\, D\, A \quad,\quad
A= (a_+,a_+^\dag,a_-,a_-^\dag)^t\quad,\quad A = S\, \mathfrak{Z}
\label{hfact}
\end{eqnarray}
where $D$ is some diagonal positive matrix, $S$ is some linear
transformation such that $(a_\pm,a_\pm^\dag)$ 
satisfy decoupled bosonic commutation 
relations $[a_\pm,a_\pm^\dag] = 1\!\! 1_q$. 

\noindent To begin with the factorization of the 
noncommutative Hamiltonian (\ref{hamat}), we introduce the vectors
\begin{eqnarray}
&& A^+ = (a_+^\dag,a_+,a_-^\dag,a_-)^t=\Lambda A\quad, \quad
{\mathfrak{Z}}^+= (\hat{\bar{Z}}',\hat{Z}',\hat{P}_{\bar{Z}},\hat{P}_{Z})^t=
\Lambda \mathfrak{Z}  
\end{eqnarray}
with the permutation matrix
\begin{eqnarray}
\Lambda = \left(\begin{array}{cccc}
0 &1&0&0\\
1&0&0&0 \\
0&0&0 &1\\
0&0&1&0 
\end{array}\right).
\end{eqnarray}
Note that $\Lambda^2 =\mathbb{I}_4$ and $(A^+)^t = A^\dag$ 
and the linear transformation $S$ is such that 
\begin{eqnarray}
 A^+ =S^* \,\mathfrak{Z}^+ 
\label{rela}
\end{eqnarray}
with $\Lambda S \Lambda = S^*$.
Another important ingredient in the factorization 
is the matrix $\mathfrak{g}$ with entries 
\begin{eqnarray}
\mathfrak{g}_{lk} =  [\mathfrak{Z}_l,\mathfrak{Z}^+_k], \quad
l,k=1,\ldots,4.
\label{gmetric}
\end{eqnarray}
A simple verification shows that $\mathfrak{g}$ is Hermitian
and therefore has the natural property to be 
diagonalizable by a matrix of different orthogonal eigenvectors 
that we shall denote by $u_i$, $i=1,\dots,4,$ 
associated with real
eigenvalues $\lambda_i$.
Now since $(a_{\pm}, a_{\pm}^{\dagger})$ satisfy the bosonic
commutation relations, we have the following 
algebraic constraints on the pair $(A,A^+)$ 
\begin{eqnarray}
[A_k, A^+_m] = (\mathbb{J}_{4})_{km}\qquad,\qquad
\mathbb{J}_{4} = {\rm diag}(\sigma_3,\sigma_3)\qquad,\qquad
\sigma_3 =
\left(\begin{array}{cc}
1&0\\
0&-1\end{array}
\right)
\label{defn}
\end{eqnarray}
which leads to the key identity
\begin{eqnarray}
S \, \mathfrak{g}\, S^\dag = \mathbb{J}_{4}.
\label{key}
\end{eqnarray} 
Another important property of $\mathfrak{g}$ 
which simplifies thing further reads
\begin{eqnarray}
\Lambda\, \mathfrak{g}\, \Lambda = -\mathfrak{g}^*
\label{lamglam}
\end{eqnarray}
from which it can be shown that if $u_i$ 
is an eigenvector of $\mathfrak{g}$ associated with
the eigenvalue $\lambda_i$, then $\Lambda u_i^*$ 
is also an eigenvector of $\mathfrak{g}$ 
associated with the eigenvalue $-\lambda_i$. 
From this property, the structure of the eigenbasis 
of $\mathfrak{g}$ follows to be $\{u_1, \Lambda u_1^*, 
u_2, \Lambda u_2^*\}$ and its associated spectrum is
$\{ \lambda_1 ,-\lambda_1 ,\lambda_2 ,-\lambda_2 \}$.

\noindent The spectral structure of $\mathfrak{g}$ 
allows us to diagonalize the Hamiltonian easily. 
We start by choosing the matrix
$S^\dag= (u_1,\Lambda u_1^*, u_2, \Lambda u_2^*)$
to be the eigenvector matrix of $\mathfrak{g}$. 
A rapid checking proves that indeed $\Lambda S \Lambda = S^*$.
Furthermore from (\ref{key}) we have 
$(S^\dag)^{-1}=  \mathbb{J}_{4} \,S\,\mathfrak{g}$ and 
$S^{-1} = \mathfrak{g}\, S^\dag \mathbb{J}_{4}$
and from (\ref{hfact}) and (\ref{rela}) follow  
$\mathfrak{Z}=S^{-1}A$ and $\mathfrak{Z}^{+}=(S^*)^{-1}A$.
Substituting these relations in (\ref{hamat}), we obtain
the new operator 
\begin{eqnarray}
\hat{H}= \frac{1}{4m} \; (A^+)^t \,\mathbb{J}_{4}\,S\, 
\mathfrak{g}^2\, S^\dag \,\mathbb{J}_{4}\, A.
\label{HAM}
\end{eqnarray} 
It should be noted that the eigenvectors
have not yet been normalized. 
The normalization conditions can be fixed such that,
for $ i=1,2$, we have
\begin{eqnarray}
(u_i)^\dag u_i =1/ |\lambda_i|
\label{normc}
\end{eqnarray} 
which also guarantees the positivity of the diagonal matrix $D$.

\noindent This finally leads to the diagonalized Hamiltonian 
\begin{equation}
\hat{H}= \frac{1}{4m} \left\{  \lambda_+(a^\dag_+ a_+ + a_+ a^\dag_+) 
+ \lambda_- (a^\dag_- a_- + a_- a^\dag_-) \right\} 
\label{diagham}
\end{equation} 
where $\lambda_\pm$ are the positive 
eigenvalues of the matrix $\mathfrak{g}$, which reads
\begin{equation} 
\mathfrak{g}=
\left(\begin{array}{rcllll}
\mathbf{a}_\theta&0&0& i{\mathbf{b}}_{\tilde\hbar}\\
0&-\mathbf{a}_\theta&i{\mathbf{b}}
_{\tilde\hbar}&0\\
0&-i{\mathbf{b}}_{\tilde\hbar}&2 \bar\theta&0\\
-i{\mathbf{b}}_{\tilde\hbar}&0&0&-2 \bar\theta\\
\end{array}\right) \,
\label{gg}
\end{equation}
where ${\mathbf{a}}_\theta=2 m^2\omega^2\theta$ and 
${\mathbf{b}}_{\tilde\hbar}=2\, \tilde\hbar\,m\omega$. The 
eigenvalues of $\mathfrak{g}$ can be easily determined and read
\begin{equation} 
\lambda_{+} = \frac{1}{2}\left(-2 \bar\theta + {\mathbf{a}}_\theta + \sqrt{\Delta_{\theta,\tilde\hbar,\bar\theta}}\right)\;\;,\;\;
\lambda_{-} = \frac{1}{2}\,\left(2 \bar\theta -{\mathbf{a}}_\theta + \sqrt{\Delta_{\theta,\tilde\hbar,\bar\theta}}\right)\;\;,\;\;
\Delta_{\theta,\tilde\hbar,\bar\theta} =  (2 \bar\theta +{\mathbf{a}}_\theta)^2+ 4 {\mathbf{b}}_{\tilde\hbar}^2
\end{equation}
thereby achieving the diagonalization of the Hamiltonian 
(\ref{hamat}).
This operator is nothing but the sum of a pair of
one-dimensional harmonic oscillator Hamiltonians
for each coordinate direction with
different frequencies both encoding noncommutative 
parameters and deformed Planck's constant $(\theta,\tilde\hbar,\bar\theta)$.
Using the Fock space basis $|n_+,n_-\rangle$, 
one gets the energy spectrum of (\ref{hamat})
\begin{eqnarray}
E_{n_+,n_-}  = \frac{1}{2m}
 \left\{  \frac{1}{2}(-2 \bar\theta + {\mathbf{a}}_\theta + \sqrt{\Delta_{\theta,\tilde\hbar,\bar\theta}}) n_+ 
 + \frac{1}{2}\,(2 \bar\theta -{\mathbf{a}}_\theta + \sqrt{\Delta_{\theta,\tilde\hbar,\bar\theta}}) n_- +  \sqrt{\Delta_{\theta,\tilde\hbar,\bar\theta}} \right\}.
\label{eigen}
\end{eqnarray} 
Setting $\bar\theta=0$, we get the spectrum of \cite{sj7} 
computed by the same diagonalization technique.
This also agrees with the result of \cite{mitra} obtained via Bopp shift representation
without momentum-momentum noncommutativity.

\noindent We now make another interesting 
observation. The energy spectrum (\ref{eigen}) 
shows that the time reversal symmetry
is broken as expected for any noncommutative theory. Nevertheless, we find that there exists a particular 
choice of $\bar\theta$ for which the time reversal symmetry    gets restored. Indeed, setting $\mp 2 \bar\theta \pm {\mathbf{a}}_\theta =0$,
equivalently $\bar\theta = m^2\omega^2\theta$, 
we get a degenerate energy spectrum.  
This result is in fact
impossible to obtain in a quantum phase-space 
where we have only space-space noncommutativity.

\noindent Finally, we consider the problem of a charged particle on a noncommutative plane subjected to 
a magnetic field $\vec{B}= B\hat{k}$ 
($\hat{k}$ is the unit vector along the 
$z$-direction) orthogonal to the plane 
along with a harmonic oscillator potential of frequency $\omega$. In the symmetric gauge 
$\vec{A}(\vec{X})=(1/2)\vec{X}\times \vec{B}$, $\vec{X}=(\hat{X},\hat{Y})$, the Hamiltonian reads
\begin{eqnarray}
\hat{H}=\frac{1}{2m}
\left(\hat{P}_{i}+\frac{eB}{2}\epsilon_{ij}\hat{X}_j\right)^2 + \frac{1}{2}m\omega^2
\hat{X_i}^2~.
\label{eigen1a}
\end{eqnarray} 
We now show that the above method can be applied to diagonalize an even more general noncommutative Hamiltonian
of this form. 
Using the algebra (\ref{alg1ab1}), we obtain the following algebra
between $\hat{X}_i$ and the canonically conjugate
momenta $\hat{\Pi}_{j}=\hat{P}_{i}+(eB/2)
\epsilon_{ij}\hat{X}_j$ :
\begin{eqnarray}
[\hat{X}_i, \hat{X}_j]=i\theta\epsilon_{ij}\quad,
\quad [\hat{X}_i, \hat{\Pi}_j]= i \tilde{\tilde{\hbar}}\delta_{ij}
\quad,\quad[\hat{\Pi}_i, \hat{\Pi}_j]=-i\bar{\bar\theta}\epsilon_{ij}
\label{alg1aa}
\end{eqnarray} 
where $\tilde{\tilde{\hbar}}=\tilde{\hbar} + eB\theta/2$ and 
$\bar{\bar\theta}=\bar\theta - eB\tilde{\hbar} - e^2 B^2\theta/4$.
 
\noindent The complex operators $\hat{Z},\hat{\bar Z} $
(as defined earlier) and $\hat{\Pi}_{Z}=\hat{\Pi}_{X}-i\hat{\Pi}_{Y}$,
$\hat{\Pi}_{\bar Z}=\hat{\Pi}_{X}+i\hat{\Pi}_{Y}$ can then be used
to recast the Hamiltonian (\ref{eigen1a}) in the form 
(\ref{hamat}) with 
$\mathfrak{Z}=(\hat{Z},\hat{\bar Z}, \hat{\Pi}_{Z}, \hat{\Pi}_{\bar Z})$ 
and they satisfy the algebra
\begin{eqnarray} 
[\hat{Z},  \hat{\bar Z} ] = 2\theta\quad,\quad
[\hat{Z}, \hat{\Pi}_{Z} ] = 2i \tilde{\tilde{\hbar}} =
[\hat{\bar Z}, \hat{\Pi}_{\bar Z} ]\quad,\quad
[\hat{\Pi}_{Z}, \hat{\Pi}_{\bar Z} ]= 2\bar{\bar\theta}.
\label{qbalg2aa}
\end{eqnarray} 
One can now easily obtain the 
matrix $\mathfrak{g}$ (\ref{gg}) by replacing
$\tilde{\hbar} \to \tilde{\tilde{\hbar}}$ and $\bar\theta\to \bar{\bar\theta}$ in $\mathbf{b}_{\tilde{\hbar}}$
and $\Delta_{\theta,\tilde{\hbar},\bar\theta}$
which finally leads to the energy spectrum 
\begin{eqnarray}
E_{n_+,n_-}  = \frac{1}{2m}
 \left\{  \frac{1}{2}(-2 \bar{\bar\theta} + {\mathbf{a}}_\theta + \sqrt{\Delta_{\theta,\tilde{\tilde\hbar},\bar{\bar\theta}}}) n_+ 
 + \frac{1}{2}\,(2 \bar{\bar\theta} -{\mathbf{a}}_\theta + \sqrt{\Delta_{\theta,\tilde{\tilde\hbar},\bar{\bar\theta}}}) n_- +  \sqrt{\Delta_{\theta,\tilde{\tilde\hbar},\bar{\bar\theta}}} \right\}.
\label{eigenspec}
\end{eqnarray} 
Once again, we observe that there exists a choice of
$\bar\theta$ given by
\begin{eqnarray}
\bar\theta= \left(m^2\omega^2 + \frac{e^2 B^2}{4}\right) \theta+ e B \tilde{\hbar}
\label{choice} 
\end{eqnarray}
for which the time reversal symmetry is restored.

\section*{Acknowledgments}
This work was supported under a grant of the  National Research Foundation of South Africa.

\end{document}